# GQM+Strategies®: A Comprehensive Methodology for Aligning Business Strategies with Software Measurement

*Victor R. Basili* [A, D], *Jens Heidrich* [B, C], *Mikael Lindvall* [A],
*Jürgen Münch* [B], *Myrna Regardie* [A], *Dieter Rombach* [B, C],
*Carolyn Seaman* [A, D], *and Adam Trendowicz* [B]

[A] Fraunhofer CESE, [B] Fraunhofer IESE, [C] University of Kaiserslautern,
and [D] University of Maryland

basili@fc-md.umd.edu, jens.heidrich@iese.fraunhofer.de, mikli@fc-md.umd.edu,
juergen.muench@iese.fraunhofer.de, mregardie@fc-md.umd.edu,
dieter.rombach@iese.fraunhofer.de, cseaman@fc-md.umd.edu,
adam.trendowicz@iese.fraunhofer.de

*Abstract:*

*In software-intensive organizations, an organizational management system will not guarantee organizational success unless the business strategy can be translated into a set of operational software goals. The Goal Question Metric (GQM) approach has proven itself useful in a variety of industrial settings to support quantitative software project management. However, it does not address linking software measurement goals to higher-level goals of the organization in which the software is being developed. This linkage is important, as it helps to justify software measurement efforts and allows measurement data to contribute to higher-level decisions. In this paper, we propose a GQM+Strategies® measurement approach that builds on the GQM approach to plan and implement software measurement. GQM+Strategies® provides mechanisms for explicitly linking software measurement goals to higher-level goals for the software organization, and further to goals and strategies at the level of the entire business. The proposed method is illustrated in the context of an example application of the method.*

## 1  Introduction and Motivation

GQM+Strategies® is a fairly new measurement approach based on the familiar Goal Question Metric paradigm [1], which is today in widespread use for creating and establishing measurement programs throughout the software industry. This new extension to GQM adds the capability to create measurement programs that ensure alignment between business goals and strategies, software-specific goals, and measurement goals.

---

® Registered trademark application pending, Fraunhofer USA and Fraunhofer IESE.

GQM has served the software industry well for several decades, in defining their measurement programs. Yet, it does not provide explicit support for integrating its software measurement model with elements of the larger organization, such as higher-level business goals, strategies, and assumptions. This linkage is important for several reasons. Software engineers are, for instance, frequently faced with apparently unrealistic goals related to software development. For example, if the next version of a product with some embedded software needs to be released to the market in half of the originally planned time, the software development schedule is often simply cut in half. There is rarely a discussion of trade-offs or other options to avoid unsatisfactory results. Specific business goals and strategies need to be defined explicitly and derived from business goals in a systematic and transparent way.

Another aspect of the problem is that software improvement strategies, such as CMMI [4] and ITIL [8], are not directly linked to generating business value. Thus, it is not sufficient to merely follow such strategies; the strategies need to be linked to business goals. In practice, an approach to measurement is needed that explicitly links high-level business goals and software measurement data.

## 2       Related Work

Several structured approaches to software measurement have been developed and are used in organizations. These approaches are referred to as "goal-oriented" approaches because they use goals, objectives, strategies, or other mechanisms to guide the choice of data to collect and analyze in a systematic way. One well-known goal-oriented approach is the GQM approach [1]. GQM provides a method for an organization or a project to define goals, refine those goals down to specifications of data to be collected, and then analyze and interpret the resulting data with respect to the original goals. Balanced Scorecard (BSC) [6] is an approach to linking measurement at all levels of an organization to the organization's strategy. The "scorecard" consists of four perspectives: financial, customer, internal business processes, and learning and growth. BSC does not dictate a static set of measures, but serves as a framework for choosing measures, processes, and initiatives that are aligned with organizational strategy and higher-level business goals. Practical Software Measurement (PSM) [10] offers very detailed guidance on software measurement, including a catalogue of specific measures, along with information on operationalizing them in an organization. PSM also includes a process for choosing appropriate measures based on the issues and objectives relevant to a software development project.

[2] addresses the misalignment between strategy at the organizational and project levels of software organizations caused by project data that does not address organizational goals and organizational goals that fail because they are not opera-

tionalized through processes and metrics at the project level. Their approach is to embed a GQM structure within each of the four BSC perspectives. One advantage of this approach is that it allows more consistency in notation and terminology surrounding goals and measures at all levels. However, the approach does not recognize or support truly different goals at different levels of the organization. The M3P – Model, Measure, Manage Paradigm – is an extension of the QIP and GQM [7]. M3P embeds GQM as a measurement definition technique within a larger framework that encompasses organizational concerns. M3P does not allow for goals at different levels of the organization that are different but explicitly linked, to allow analysis of measurement data to feed back up the chain.

There is an increasing awareness that the IT infrastructure supporting business processes of a company imposes significant risks for the company. As a consequence, several regulatory constraints in the IT governance domain and the IT service domain (such as Sarbanes Oxley, SOX) emphasize the need for linking business goals to properties of the IT infrastructure (such as availability). The solutions proposed by models in these domains (especially COBIT® 4.1 [5] and ITIL release 3) only offer simple connections between predefined sets of goals and attributes of the IT infrastructure. Such models are not focusing on risks that are imposed in the case that software is a product or software is part of a product of a company. In addition, they do not support specifying individual goals including all dimensions. GQM⁺Strategies® can focus on all software-related activities and aspects of a company and integrate them into its business strategies.

## 3    The GQM Approach

The GQM approach [1] develops a set of so-called measurement goals for aspects of the software product and process of interest (e.g., productivity and quality), derives questions (based upon models) that define those goals as completely as possible in a quantifiable way, specifies the measures that need to be collected to answer those questions, and tracks process and product conformance to the goals. A GQM measurement goal is defined using five basic terms described in Table 1.

Measurement goals may be defined for any object, for a variety of reasons, with respect to various models of quality, from various points of view, and relative to a particular environment. Such goals are refined into specific questions that must be answered in order to evaluate the achievement of the goal. The questions are then operationalized into specific quantitative measures. All the components of the goal are used in formulating the questions and specifying the measures by narrowing down the space of inquiry to those questions and metrics that are relevant to the goal and also by making sure all aspects of the goal are covered by the questions and metrics.

In order to create a comprehensive measurement plan, mechanisms for data col-lection have to be developed and an interpretation model has to be defined. These are models of the products, processes, and quality perspectives of interest from which the metrics themselves are defined. Interpretation models, as the name implies, help practitioners interpret the data yielded by the metrics.

After that, the measurement program can be applied to collect, validate, and analyze the data in real time in order to provide feedback to projects, e.g., for initiating corrective actions. Finally, the collected data can be analyzed in a postmortem fashion to assess conformance to the goals and make recommendations for future improvements.

|  | Description | Example |
|---|---|---|
| **Object** | process, product, other experience model | Analyze the system test process |
| **Purpose** | characterize, evaluate, predict, motivate, improve | for the purpose of evaluation |
| **Focus** | cost, correctness, defect removal, changes, reliability, user friendliness, ... | with respect to defect slippage |
| **Viewpoint** | user, customer, manager, developer, corporation, ... | from the point of view of the corporation |
| **Context** | problem factors, people factors, resource factors, process factors, ... | in the context of organization X |

**Table 1:** Template for defining GQM measurement goals.

GQM was originally developed in the NASA Goddard Space Flight Center's flight software development environment, which was also the originating environment for the Quality Improvement Paradigm (QIP). The QIP is an evolutionary model tailored for improvement of software development organizations. It includes a goal-setting step, as well as steps that involve measurement and evaluation of improvement efforts. GQM is used in these steps of the QIP, so the two concepts are closely related.

## 4      The GQM$^+$Strategies$^®$ Approach

The GQM$^+$Strategies$^®$ method adds several extensions on top of the GQM model in order to make the business goals, strategies, and lower-level goals explicit. Strategies are formulated to deal with business goals such as improving customer satisfaction, garnering market share, reducing production costs, taking into account the context and making explicit any assumptions. Strategies help define lower-level goals that can be assigned to different parts of the organization, e.g., software-specific goals, hardware goals, marketing goals, etc. GQM$^+$Strategies$^®$ also makes the relationships between concrete activities and measurement goals explicit. Sequences of activities necessary for accomplishing the goals are defined by the organization and embedded into strategies in order to achieve some goal. Links are established between each goal and the strategy it supports. Attached to goals, strategies, and scenarios at each level of the model is information about relationships between goals, relevant context factors, and assumptions.

The entire model provides an organization with a mechanism not only to define measurement consistent with larger, upper-level organizational concerns, but also to interpret and roll up the resulting measurement data at each level. GQM⁺Strategies® linkages and measures ensure that the business goals are fulfilled.

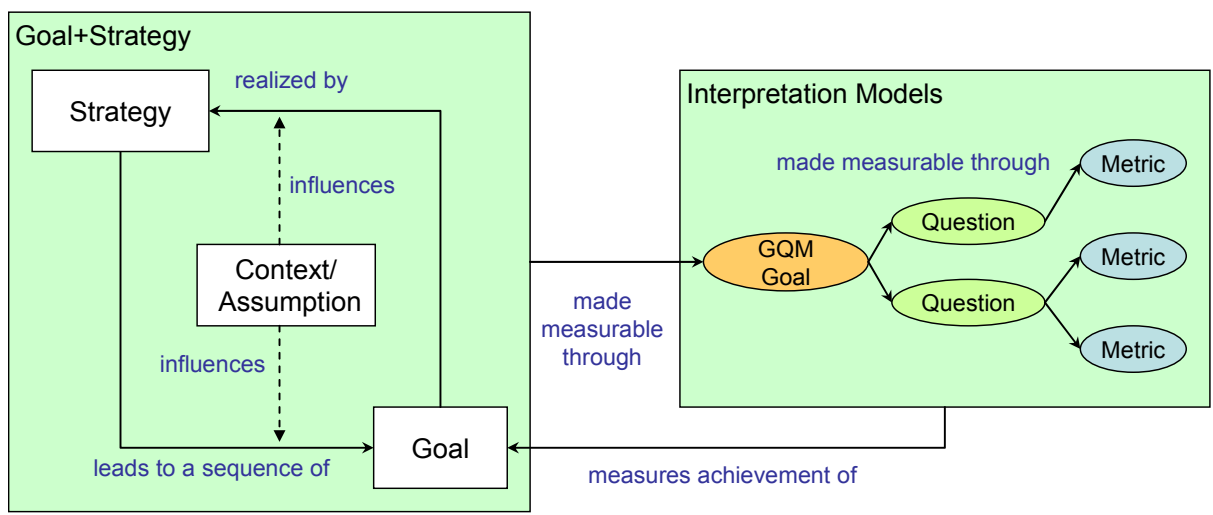

**Figure 1:** Components of the GQM⁺Strategies® meta-model.

Figure 1 gives an overview of all conceptual components for constructing a consistent model. The GQM⁺Strategies® meta-model allows multiple goal levels and permits deriving multiple strategies for each of these goal levels. A goal may be realized by a set of concrete strategies, which may in turn lead to a sequence of goals. A set of predefined goals and strategies may be defined as part of an (organization-specific) experience base. Context information about the organization and assumptions are drivers for this instantiation process and influence the definition of new goals and strategies as well as the selection and adaptation process of predefined goals and strategies.

At each level of the instantiation process, GQM plans are defined in order the measure the achievement of the defined goal in combination with the chosen strategy. This includes the definition of GQM measurement goals, the derivation of questions and metrics, as well as the definition of an interpretation model that determines whether the measurement goal has been reached. Table 2 gives an overview of the basic GQM⁺Strategies® terminology.

| **Business goals** | Goals the organization wishes to accomplish in general in order to achieve its objectives. |
|---|---|
| **Context factors** | Environmental variables that represent the organizational environment and affect the kind of models and data that can be used |
| **Assumptions** | Estimated unknowns that can affect the interpretation of the data. |
| **Strategy decisions** | A set of possible approaches for achieving a goal. |
| **Strategy activities** | A set of activities to help achieve the chosen strategy. |
| **Lower Level Goals** | A set of goals inherited from upper-level goals as part of the upper-level goal strategy, e.g., a goal related to the software process or product for implementing system strategy decision. |
| **GQM Goals** | Goals defined so that they can be measured using the GQM approach. A GQM goal is associated with goals at all levels. |
| **Interpretation models** | Models that help interpret data to determine whether goals at all levels are achieved. |

**Table 2:** Basic terminology of the GQM$^+$Strategies$^®$ approach.

Four different types of business goals are distinguished: growth goals, success goals, maintenance goals, and specific focus goals. Growth goals include acquiring new projects within the current competence areas, expanding the existing project set, evolving existing competencies, building new competencies, etc. Success goals include delivering good products to customers, controlling costs, shrinking schedules, increasing profits, achieving corporate visibility (e.g., awards), or building core competencies. Maintenance (internal) goals include transparency, employee satisfaction, controlled risk, learning environment; the idea is to measure to assure no erosion of the status quo. Specific focus goals include such things as making the helpdesk more efficient, or predicting if a proposed effort has a good ROI.

The GQM$^+$Strategies$^®$ method consists of the following basic tasks for defining a corresponding model (meta-model instantiation):

*Task 1: Determine and define business goals.* In order to define a business goal for an organization, the basic motivation needs to be determined and the business goal and the rationale that lead to defining the goal (making use of context factors and assumptions) need to be described. For instance, one wants to safeguard a place in the market and therefore increases customer satisfaction. After that, the business goal needs to be formalized, which is facilitated by filling out a goal template. A description of this template and examples are presented in Table 3.

|  | Description | Examples |
|---|---|---|
| **Activity** | The basic activity that should be performed in order to accomplish the goal. | Reducing, increasing, achieving, pursuing, or providing the main focus of the business goal. |
| **Focus** | The main (quality) focus of the business goal. | Cost, profit, turnover, market share, prestige, customer satisfaction. |
| **Object** | The object under consideration. | People, market, a project, collection of projects, customer. |
| **Magnitude (degree)** | The quantification of the goal specified by a magnitude. | 10%, $1M, 5% more than last year. |
| **Timeframe** | The timeframe in which the magnitude has to be achieved. | 3 years, up till January 1, 2008, or permanently. |
| **Scope** | The scope of the goal. | The whole organization, a certain business unit, or a person. |
| **Constraints (limitations)** | Basic constraints that may limit accomplishing the goal. | Limited influence on certain factors, laws, mission statement, and basic principles. |
| **Relations with other goals** | Potential relations with other (complementary or competing) goals. | Tradeoffs, hierarchy, and ordering. |

**Table 3:** Template for defining GQM⁺Strategies® goals.

*Task 2: Select strategy decisions for business goals.* A list of potential strategies for achieving the business goal needs to be identified (e.g., building a more reliable system that will lead to less customer complaints vs. testing reliability in). Then, the most promising strategy has to be selected considering feasibility, cost, and benefit, and taking into account context factors and assumptions that naturally restrict the set of applicable strategies.

*Task 3: Develop a measurement plan for business goals and strategy decisions.* A measurement plan is developed for evaluating the achievement of the business goal (from task 1) and the corresponding strategy decision (from task 2). Therefore, potential measurement goals need to be identified. The GQM goal template (object, purpose, quality aspect, viewpoint, and context) is used for formalizing the measurement goal. After that, GQM questions and metrics and criteria for evaluating the achievement of the measurement goals are determined and interpretation models are defined for aggregating and interpreting the collected measurement data. With that, the GQM⁺Strategies® model provides guidance not just for planning, but also for analyzing and rolling up the resulting data to the decision makers and helps to make the right decisions for achieving the designated business goal and evaluating the implementation strategy. It does so by helping manage expectations throughout the process by providing visibility into measures such as ROI (which is a good thing especially for a business making a substantial financial investment into a major process improvement initiative).

*Task 4: Determine and define lower-level goals.* The implications of the chosen upper-level strategies (e.g., strategy decision of the business level) with respect to lower-level goals (e.g., software development-specific goals) have to be elicited. For instance, if a chosen strategy is to test in reliability, the software test processes must be examined. Potential lower-level goals need to be identified based on this analysis (e.g., decrease customer reported defects by improving system test effectiveness). The most promising goal considering feasibility, cost, and benefit is then selected in order to break down the upper-level strategy decisions. Again, context factors and assumptions help to define the right selection criteria. After that, the lower-level goal needs to be formalized using the goal definition template (as described for defining business goals in task 1).

*Task 5: Select strategy decisions for lower-level goal.* Based upon the lower-level goals, a list of potential strategy decisions is identified. Each strategy decision may be further refined by a set of strategy-related activities that may be executed or not, depending on a list of context factors and assumptions addressing questions such as: Does historical data related to my goal and interpretation model exist; are the projects from the historical data set relevant; or are experts available to make intelligent estimates for the historical baselines? Then, the most promising strategy has to be selected.

*Task 6: Develop a measurement plan for lower-level goals and strategy decisions.* Again, a measurement plan is developed for evaluating the achievement of the lower-level goal (from task 4) and the corresponding strategy decision (from task 5). Measurement goals need to be identified, questions and metrics need to be derived, and an interpretation model needs to be defined in order to make statements about the achievement of the goal and the success of the applied strategy.

Tasks 3 to 6 may be iterated in order to further break down the goal and strategy hierarchy. The stop criterion depends upon the organization for which the GQM$^+$Strategies$^®$ model is built. Usually, the process stops if a concrete set of steps has been derived that may easily be implemented in the organization (e.g., on the project level).

Developing a GQM plan for each goal and strategy level is not an isolated task; that is, the metrics derived across different levels of the GQM$^+$Strategies$^®$ model will usually overlap. Also, an interpretation model for a higher-level goal may only be defined completely if the lower-level goals have been already modeled. This implies that there is no strict sequence for developing a GQM$^+$Strategies$^®$ model. One may do things in parallel, bottom-up, top-down, or start somewhere in the middle.

Multiple goals on the same abstraction level or on different levels of the GQM⁺Strategies® model will most likely be interrelated with each other. The most obvious relationship (documented in the conceptual model) is a hierarchical structure consisting of a top goal and sub-goals connected via strategies. For a certain goal, a set of complementary goals may exist, which provide additional support for the current goal. On the other hand, a set of competing goals may also exist that conflict with the current goal, whereas other goals may be totally unaffected by the current goal.

## 5     Comprehensive Example

In the following section, a comprehensive example illustrating the method will be presented using the concrete tasks for defining the GQM⁺Strategies® model described in the previous section. The overall model consists of 3 levels described in separate sub-sections.

### 5.1     Level 1 Goal: Increase Profit

*Determine and define business goals*: Organization ABC provides information services to customers through the Web. Customers pay for accessing information via the software that searches, analyzes, and presents information, not for the software itself. The amount of revenue generated at ABC is determined by the number of times customers access the ABC software products and use the services provided. The ABC business goal is to increase profit by 15% per year beginning in 2 years. Table 4 presents a formalization of this business goal using the GQM⁺Strategies® goal template.

| **Activity** | Increase |
|---|---|
| **Focus** | Profit |
| **Object** | ABC web-service business |
| **Magnitude (degree)** | 15% per year |
| **Timeframe** | Annually, beginning in 2 years |
| **Scope** | All divisions of the ABC web-service business |
| **Constraints (limitations)** | Available resources, … |
| **Relations with other goals** | Maintain product quality, … |

Table 4:     ABC level 1 goal: increase profit.

*Select strategy decisions for business goals*: A list of potential strategies for the business goal would be, for instance, (a) to deliver added functionality at regular and frequent intervals, (b) to increase the rates charged to customers, or (c) to reduce development costs. An assumption made by ABC was that added functionality will encourage more usage and consequently increase profit. However, this can only be true if the development costs and schedule are under control, which is the case at ABC as previous experience had shown. Consequently, strategy decision (a) was chosen.

*Develop a measurement plan for business goals and strategy decisions*: Finally, a GQM plan including an interpretation model is defined in order to evaluate the business goal for the chosen strategy decision. The GQM measurement goal can be phrased as follows: Analyze the trend in profit for the purpose of evaluation with respect to a 15% increase in annual profit per year from the point of view of ABC's management in the context of the ABC web-service business. Based on this goal, we can derive the following questions: What is the profit figure for this year $P_0$? What is the expected profit figure for each succeeding year $P_x$?

The resulting (simplified) interpretation model is as follows: Starting in year 2, if $P_x > 1.15\ P_{x-1}$, then the goal has been satisfied; else, if added functionality was increased appropriately, then some assumption or the strategy chosen was wrong. However, the full interpretation is also dependent on the lower-level goals, because if those are not reached, the chosen strategy at the top level may still have been the right one.

### 5.2 Level 2 Goal: Deliver Increased Functionality

*Determine and define lower-level goals*: Based upon the chosen level 1 (business) strategy, the next goal level is defined: Deliver 5% more requirements compared to the prior release for each project of the ABC web-service business every 6 months. That is, ABC made the assumption (based on their experience) that a 5% increase in new functionality will be sufficient to get a 15% increase in profit. Table 5 presents a formalization of this goal using the GQM$^+$Strategies$^®$ goal template.

| **Activity** | Deliver |
|---|---|
| **Focus** | Usable functionality |
| **Object** | Backlog of customer-requested requirements |
| **Magnitude (degree)** | 5 % more than the prior release |
| **Timeframe** | Every 6 months, beginning in 2 years |
| **Scope** | ABC web-service development groups |
| **Constraints (limitations)** | Available resources, ability to control cost and schedule for a release, … |
| **Relations with other goals** | Achievement of cost and schedule estimate accuracy, … |

Table 5: ABC level 2 goal: deliver increased functionality.

*Select strategy decisions for lower-level goal*: For implementation, ABC decided to use the MoSCoW approach [9] to determine what capabilities to deliver. MoS-CoW is a method for negotiating with the customer on the importance of delivery of each functional requirement. MoSCoW stands for: M - MUST have this, S - SHOULD have this if at all possible, C - COULD have this if it does not affect anything else, W - WON'T have this time but WOULD like in the future. ABC had experts available who could tailor, teach, and apply MoSCoW. Furthermore, ABC made the assumption that the backlog of customer-requested requirements continues to grow and requirements are characterized by M, S, C, W and the complexity of their implementation.

*Develop a measurement plan for lower-level goals and strategy decisions*: In order to evaluate the defined level 2 goal, ABC defined the following GQM goal: Analyze each 6-month release for the purpose of evaluation with respect to a 5% new functionality growth as compared to prior functionality growth from the point of view of the services project manager in the context of ABC services. In order to measure this goal, ABC asked the following questions: What was the amount of functionality delivered at each release? What was the percentage of new M, S, C, and W requirements released? What is the % growth from the prior release?

The resulting interpretation model was as follows: If at each 6-month milestone, the growth in functionality of a release (percentage of new M requirements) is greater than 5%, then the level 2 goal is satisfied for this release; else, assumptions about MoSCoW were wrong. If the level 1 goal (business goal, increase profits) is satisfied but the level 2 goal is not, then investigate why, e.g., delivery of some specific features alone caused the gain.

### 5.3  Level 3 Goal: Effectively Apply MoSCoW

*Determine and define lower-level goals*: The level 2 strategy depends upon being able to effectively apply the MoSCoW approach, which is in turn our lower-level goal on level 3 of the GQM⁺Strategies® model.

*Select strategy decisions for lower-level goal*: The strategy is split into a number of steps, namely to provide training for MoSCoW, pilot the approaches on a particular project, and compare results, if possible, with the currently applied approaches to elicit customer requirements. The current requirements selection process is based upon the manager's view of what to do next.

*Develop a measurement plan for lower-level goals and strategy decisions*: In order to evaluate the defined level 3 goal, ABC defined the following GQM goal: Analyze a pilot project for the purpose of evaluation with respect to MoSCoW as a method for identifying requirements from the point of view of the development manager in the context of an ABC web-service development project. In order to measure this goal, ABC asked the following questions: How well did the pilot project follow the MoSCoW process? Are the functions that have been changed in this release being used more frequently? How many new requirements are in the release? What percent of MUST requirements have been removed from the backlog? What is the cost of training for MoSCoW?

The resulting interpretation model was as follows: If the MoSCoW process was followed and functions that were changed in this release are being used more frequently, and if an appropriate number of MUST requirements could be removed from the backlog, and if the cost of training for MoSCoW was reasonable, then the level 3 goal is satisfied. If the level 2 goal (providing more requested functionality) is satisfied but the level 3 goal is not, then investigate why, e.g., the process was not followed appropriately.

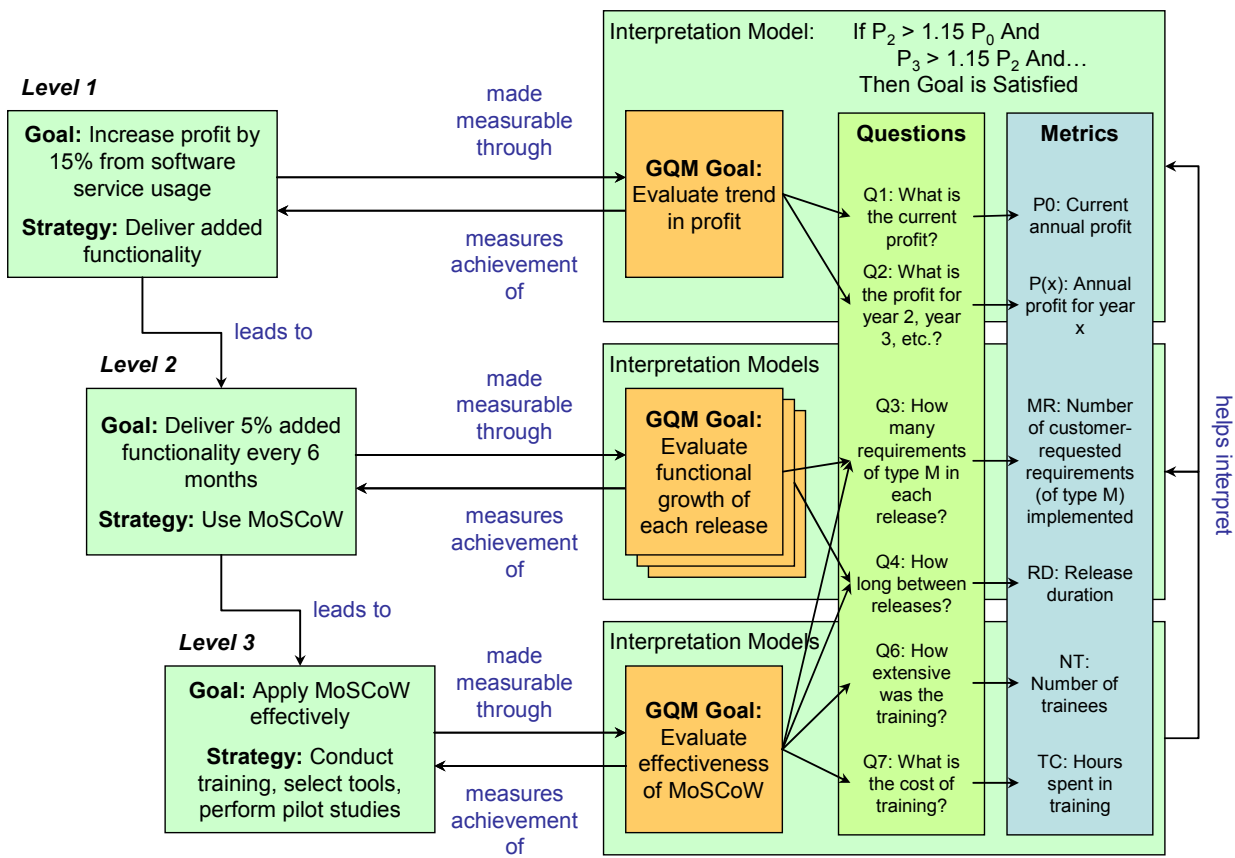

**Figure 2:** Overview of the comprehensive example.

## 5.4 Overview

Figure 2 gives an overview of the comprehensive example and illustrates the relationships between the instantiated conceptual elements of the presented model across all three abstraction levels (excluding context factors and assumptions). As indicated in the picture, GQM[+]Strategies® explicitly describes the linkages between goals at the different levels and provides interpretation models tying together measurement goals, context factors, assumptions, and concrete measurement data. This allows for transparency of the motivations for measurement and goals at different levels of the organization.

## 6  Conclusions and Future Work

In recent decades, GQM has proved to be a natural and pragmatic way of creating a measurement program based on software goals, questions, models generated by these goals, and the metrics that are needed to answer the questions. GQM⁺Strategies® inherits GQM's benefit of assuring that the metrics set is as small as possible and that the data collected address the defined organizational objectives. However, it extends GQM by providing explicit support for linking software measurement goals to organizational business objectives. Considering this linkage is essential for organizational success, as it helps to translate strategic-level objectives into a set of operational software goals and respective quantitative project management. On the other hand, it helps to justify software measurement efforts and allows measurement data to contribute to higher-level decisions. GQM⁺Strategies® provides a systematic way to deal with relationships between different objectives at various organizational levels. Early identification of conflicting objectives, for instance, may prevent failures very early, namely, at the time of defining the organizational strategy. Moreover, a transparent and intuitive way of specifying and synchronizing objectives at various operational levels contributes to better understanding of a business and to improved communication between different organizational levels.

In order to validate the practical usefulness of the GQM⁺Strategies® method, it was applied to several projects retroactively. We are in the process of building a multiple domain experience base that incorporates common business goals, strategies, specific strategy related activities, etc., and their linkages, so that software and technology organizations would be able to navigate through better the space of options and adapt it to their specific application context. The experience base is designed to support organizations in developing their own measurement programs, and in tracking and improving them over time.